# Polarized Neutron Reflectometry Study of Depth Dependent Magnetization Variation in Co Thin Film on a PMN-PT Substrate

Md Mamun Al-Rashid[1,2,*], Alexander Grutter[3,+], Brian Kirby[3,×] & Jayasimha Atulasimha[1,2,‖]

[1]Mechanical and Nuclear Engineering, Virginia Commonwealth University, Richmond, VA.
[2]Electrical and Computer Engineering, Virginia Commonwealth University, Richmond, VA.
[3]NIST Center for Neutron Research, NIST, Gaithersburg, MD.

[*]alrashidmm@vcu.edu, [+]alexander.grutter@nist.gov, [×]brian.kirby@nist.gov,
[‖]jatulasimha@vcu.edu

We studied the depth dependent magnetization profile of the magnetostrictive Co thin film layer in a PMN-PT (011)/Ta/Co/Ta structure under both zero and nonzero applied electric field using polarized neutron reflectometry. Application of electric field across the PMN-PT substrate generates a strain, which rotates the magnetization of the Co layer consistent with the Villari effect. At low magnetic fields (near remanence and coercive field conditions), we find that the depth dependent magnetization profile is non-uniform, under both zero and nonzero applied electric fields. These variations are attributable to the depth dependent strain profile in the Co film, as determined by finite element analysis simulations.

## I.  INTRODUCTION

Nanomagnetic computing, where information is encoded in the magnetization direction of magnetic nanostructures has the potential to be very energy efficient and is inherently non-volatile [1–3], and therefore amenable to novel computing architectures [4,5]. However, the energy requirements of these devices depend on the magnetization switching mechanism i.e. external magnetic field [6], spin transfer torque [7,8], spin Hall effect [9,10], voltage controlled magnetic anisotropy [11,12] or strain induced switching of magnetostrictive nanomagnets [13,14]. Among these, strain induced magnetization switching is one of the most energy-efficient techniques [13,15,16], and is also well suited to non-Boolean computing applications [17,18]. Strain clocked nanomagnetic memory and logic devices for energy efficient computing have been experimentally demonstrated by a number of groups [14,15,19–21]. These devices are typically implemented by fabricating magnetostrictive nanomagnets on top of piezoelectric substrates from which voltage induced strain is transferred to produce magnetization rotation. Such strain induced magnetization rotation has been characterized using Magneto-Optic Kerr Effect (MOKE), Magnetic Force Microscopy (MFM), Photoemission Electron Microscopy (PEEM), magnetoresistance etc. [14,15,19,22–24]. Although these techniques are excellent at resolving the average and/or near surface magnetization variation, they are unable to resolve the depth dependent magnetization profile. Variation in strain transfer from the piezoelectric substrate to the magnetostrictive layer and therefore, the depth dependent magnetization rotation of such magnetostrictive nanomagnets are yet to be studied in detail. Such variations can have important ramifications in the performance of "straintronic" nanomagnetic devices and can also lead to novel straintronic applications similar to memory devices implemented using graded media that utilizes gradually changing magnetic anisotropy normal to the film [25–27].

In this work, we examine depth dependent magnetization profile in magnetostrictive Co thin films deposited on PMN-PT (011) substrate with and without electric field induced strain using polarized neutron reflectometry (PNR), which is sensitive to the depth dependent magnetization profile [28]. We have simulated strain in such a multiferroic heterostructure using finite element analysis to estimate the depth dependent strain transfer profile from the piezoelectric substrate at different depths of the magnetostrictive film, leading to the non-uniformity in the magnetization depth profile. The paper is organized as follows. Section II describes the sample fabrication and characterization methods. Section III presents and analyzes the effect of electric field on the sample coercivity and anisotropy, COMSOL simulation of the strain transfer profile, and the PNR measurements and corresponding models of the depth dependent magnetization profile. Section IV concludes the paper. Additional modeling approaches to the PNR data is presented in the supplement.

## II. Experimental Methods

The PMN-PT substrates used in this work have a (011) orientation with lateral dimension of 10 mm × 10 mm and thickness of 0.5 mm. The single-side polished substrates have been sourced from MTI Corporation [29]. Aluminum (Al) was deposited on the unpolished side of the PMN-PT substrate to be used as the bottom electrode. A 10nm Ta layer was then deposited on the polished side as adhesion layer followed by a 60nm Co layer and 10nm Ta cap, also used as the top electrode. A schematic of the sample structure is shown Fig. 1. Vibrating sample magnetometry (VSM) measurement of the sample was performed at room temperature under zero applied voltage. PNR measurements were performed using the PBR beamline at the NIST Center for Neutron Research. The data were corrected for background, beam footprint, and imperfect beam polarization (unless otherwise mentioned). Model fitting of PNR data allows for determination of the nuclear composition and in-plane vector magnetization depth profiles of thin films and multilayers [30]. Measurements discussed here have been performed at room temperature at multiple magnetic field values, under an applied voltage of either 0 V or 400 V. Model fitting was performed using the Refl1D software [31]. Finite element analysis simulations were performed using COMSOL to estimate the strain profile in a heterostructure similar to that studied experimentally.

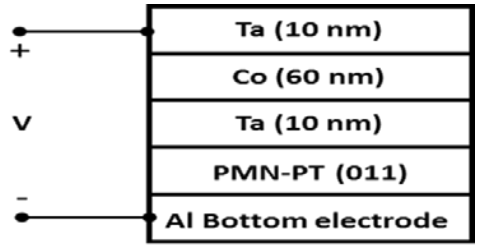

**Figure 1:** Sample schematic.

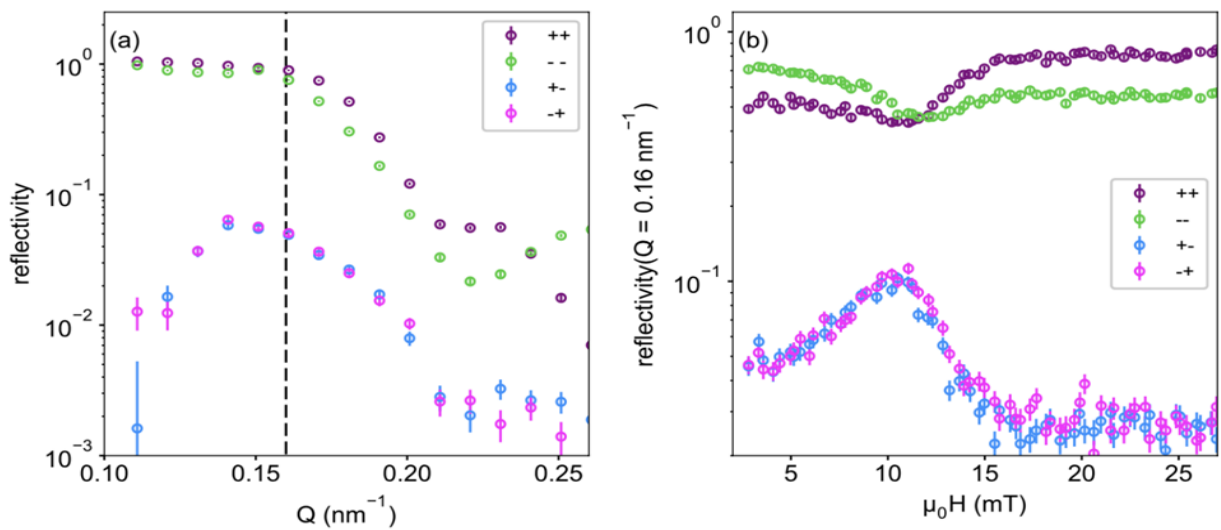

**Figure 2:** (a) Low Q PNR data measured at 1 mT and 0 V after positive magnetic saturation. (b) 0 V PNR data as function of increasing H following negative magnetic saturation. These data correspond to $Q = 0.16$ nm$^{-1}$, indicated by a dashed vertical line in (a). Error bars correspond to +/- 1 standard deviation.

## III. RESULTS AND DISCUSSION

### A. Strain Effect on Coercive Field

Accurately characterizing changes in magnetic properties as a function of electric field can be challenging using common techniques like VSM or superconducting quantum interference device (SQUID) magnetometry, due to the potential for artifacts arising from the leads. With this in mind, we used PNR to probe the voltage dependent "net" vector magnetization of our sample. Typically, a PNR measurement is carried out over a broad range of reciprocal space wavevector transfer (Q), which allows for determination of the real space depth profiles. Owing to the limited brightness of neutron sources, such measurements can be time consuming, limiting the number of conditions for a given experiment. However, the average magnetization of the entire sample can be approximated by measuring only at very low Q values, which provide sensitivity to a very broad real space length scale, and generally exhibit relatively strong reflectivities (owing to proximity to the critical edge). Figure 2 (a) shows the Q-dependent non spin-flip (++ and - -) and spin-flip (+- and -+) reflectivities at 0 V after magnetically saturating at 700 mT and returning the magnetic field to 1 mT (measurement close to remanent condition). We chose a Q value of 0.16 nm$^{-1}$, very near the critical edges, and measured the four reflectivities at that single Q point as functions of magnetic field (H) and voltage (V) after magnetically saturating in -0.7 T, as shown in Fig. 2(b).

The non spin-flip reflecitivites depend on both the nuclear and magnetic composition of a sample. Specifically, the differences in ++ and -- arise from the component of the in-plane sample magnetization parallel to H. Thus, the point where ++ and -- cross corresponds to switching of the parallel component of the magnetization from pointing opposite to H to pointing along H. The spin-flip reflectivities originate from the component of the in-plane sample magnetization perpendicular to H. As such, the peak in the spin-flip scattering shown in Fig. 2(b) corresponds to the maximum angle between the sample magnetization M and applied magnetic field H during reversal. We performed such H-dependent scans at progressively higher voltages, in order to examine the effect of electric field on the magnetic reversal process. We note that the single-Q, H scans shown are not polarization corrected. Since we are not quantitatively estimating magnetization values from these scans (only H-shifts in qualitative features), such corrections are unnecessary. Figure 3(a) shows the non spin-flip data plotted as spin asymmetry (the difference between ++ and -- divided by the sum). The electric field has a clear effect, as the center of the transition $H_0$ increases progressively by more than 1 mT with applied voltage. Here $H_0$ corresponds to the minimum in magnetization parallel to H, and is estimated by fitting the data to an error function, yielding the values shown in the Fig. 3(a) inset. A similar procedure was performed for the spin-flip data, as shown in Fig. 3(b). We expect (and observe) that +- and -+ should be identical, therefore the average spin-flip scattering is shown. In this case, $H_0$ corresponds to the maximum in the perpendicular magnetization and was determined by fitting the data to a Gaussian function, with values shown in the Fig. 3(b) inset. Again the voltage-dependent effects are significant, shifting the peak positions by more than 1 mT. Additionally, the magnitude of the spin-flip peak is observed to progressively decrease with voltage. A VSM measurement of the magnetic hysteresis loop of this sample under zero voltage is shown in Figure 4. The coercive magnetic field Hc is 12 mT, consistent with transition values shown in Fig. 3, and verifying our interpretation of the PNR signal.

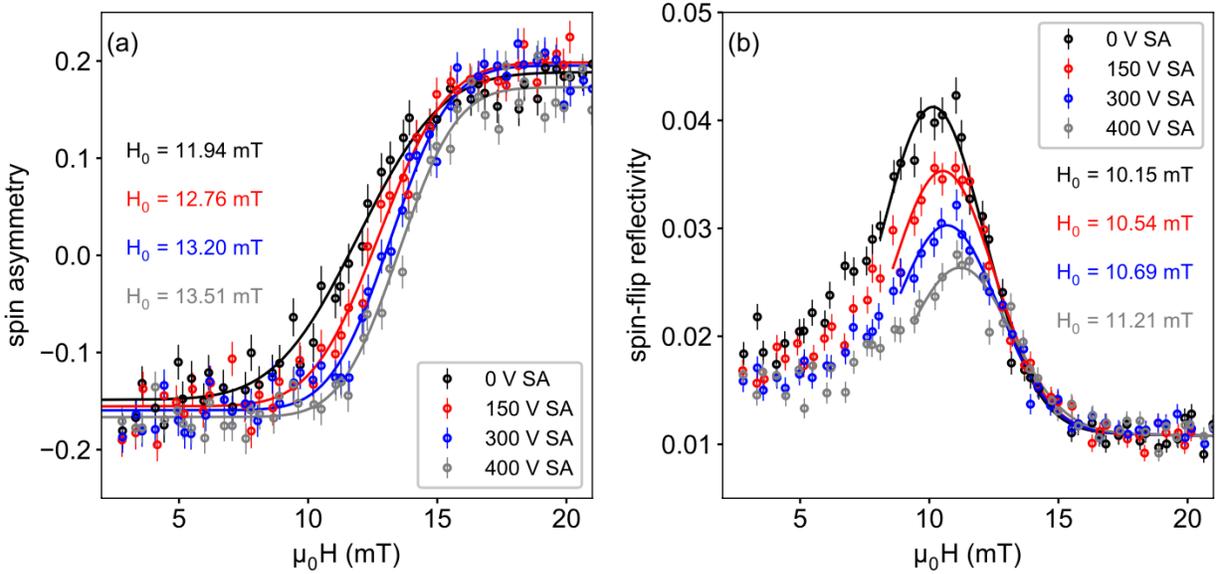

**Figure 3:** (a) Spin asymmetry, and (b) spin-flip scattering measured at $Q = 0.16$ nm$^{-1}$ as a function of increasing H following negative magnetic saturation. Measurements were repeated at progressively higher applied voltages. Error bars correspond to +/- 1 standard deviation.

The voltage-dependent shifts in magnetization minima parallel to H and corresponding shifts in magnetization maxima perpendicular to H shown in Fig. 3 suggest that the stress anisotropy for a positive applied voltage induces a magnetization "easy axis" (energetically favorable direction) along H, therefore, increasing Hc for the sample. Notably, at all voltages measured, the minima in the spin asymmetry occurs at higher H than does the maxima of the spin-flip scattering. This suggests that the magnetization reversal occurs through a combination of coherent rotation and domain nucleation as one would expect in a thin film. That the magnitude of the spin-flip peak decreases with voltage is likely a consequence of the increased anisotropy. In a system with high magnetic anisotropy, where the "easy axis" is along the applied magnetic field, it is energetically costly to rotate or reorient the magnetization through formation of domains with magnetization component perpendicular to the "easy" axis. As a result, with increasing uniaxial-anisotropy there is a higher likelihood of magnetization reversal through 180° domain formation and motion. An illustration of this is shown in Fig. 5. With increasing applied voltage and therefore stress anisotropy, domains with a perpendicular magnetization component become less energetically favorable, resulting in the observed decrease in the peak spin flip (SF) scattering reflectivities.

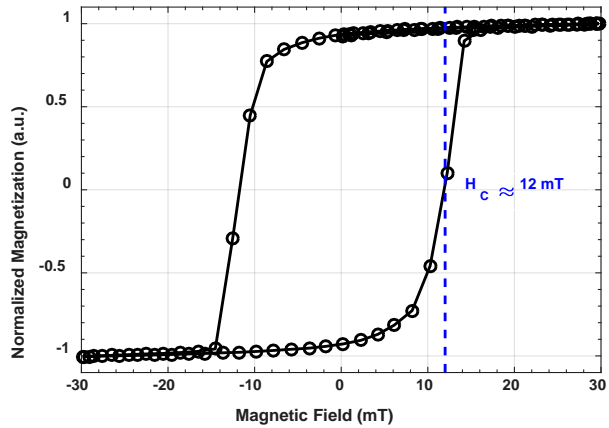

**Figure 4:** VSM measurement of the magnetic hysteresis loop at room temperature under zero voltage.

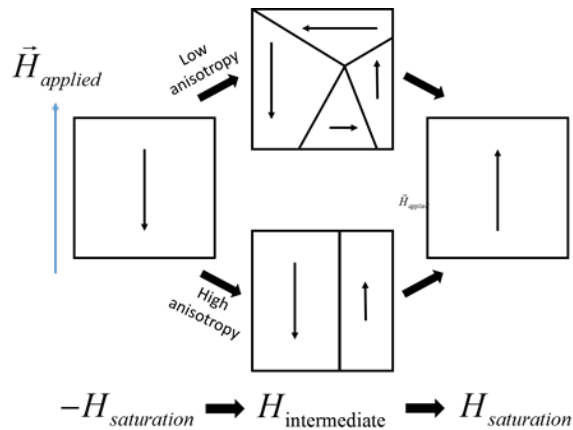

**Figure 5:** Magnetization reversal scenarios for different magnitudes of magnetic anisotropy.

## B. Depth Dependent Magnetization Rotation

Q-dependent PNR measurements were conducted at 3 different external magnetic fields – positive saturation (at 700 mT), near-remanence (at 1 mT after decreasing the field from positive saturation) and near-coercive field (10 mT after increasing the field from negative saturation) in order to determine the depth profiles. Remanence and coercive field measurements were performed at applied voltages of 0V and 400V to observe the effects of electric field induced stress on the magnetization profile. The model fitted saturation data is shown in Figure 6(a). Spin-flip scattering should not be present at saturation, thus only non spin-flip scattering was measured at this condition. The fit to the data is excellent, and corresponds to the depth profile shown in Fig. 6(b). The profiles are shown in terms of scattering length density $\rho$, which has a complex valued nuclear component (corresponding to the nuclear composition) and a magnetic component (proportional to the magnitude of the magnetization vector). The model is quite simple, with a uniform Co magnetization, and nuclear scattering length densities close to expected values for all layers. The nuclear profile determined from this fit was used for fits at all other conditions.

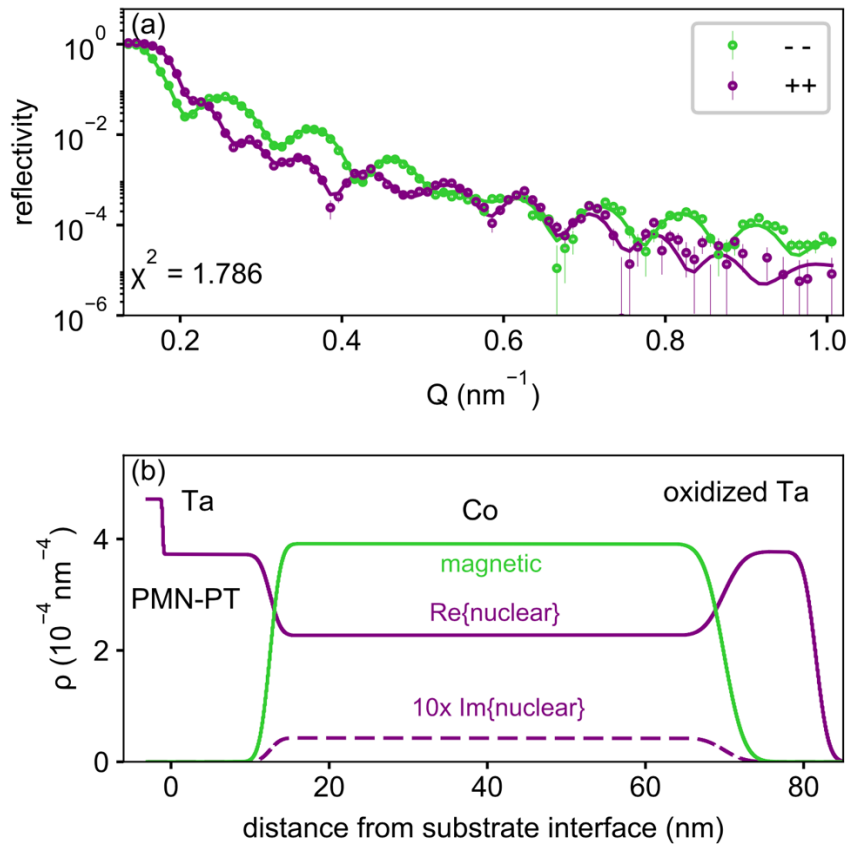

**Figure 6:** (a) Fitted PNR data taken at 700 mT and 0 V. (b) Nuclear and magnetic depth profiles used to generate the fits in (a). Note that the magnetic component of $\rho$ is proportional to the magnetization. Error bars correspond to +/- 1 standard deviation.

At lower magnetic field, we find that the data are inconsistent with a uniform Co magnetization profile. Figure 7(a) shows data taken at 1 mT in 0 V after saturating at 700 mT. Notably, there is significant spin-flip scattering, demonstrating that even at 0 V, anisotropy plays a significant role, as the magnetization relaxes away from the applied magnetic field direction. The fit in Fig. 7(b) corresponds to a profile with depth-independent magnetization magnitude and rotation angle, and does not fit the low-Q spin-flip data well. Uniform profile fits produce similarly poor results for other measurements at 1 mT and at 10 mT. We found that significant non-uniformity was required to achieve a good fit to the data, but that there were multiple non-uniform models that could provide essentially equivalent fits. To guide the modeling, we performed a COMSOL simulation of strain transfer in a similar structure. Specifically, we simulated a PZT/Ti$_{10nm}$/Co$_{60nm}$/Ti$_{10nm}$ heterostructure with Al as the bottom electrode as shown in Fig. 8(a). We have used PZT instead of PMN-PT and Ti instead of Ta in because those material parameters were readily available in the COMSOL package, but these would not change the qualitative features of the simulation of depth dependent strain transfer.

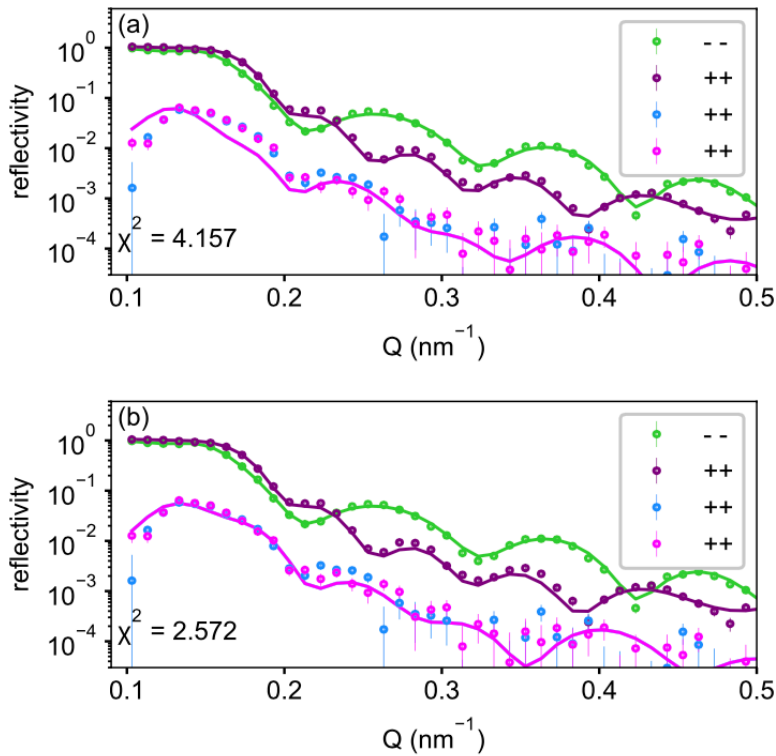

**Figure 7:** (a) 1mT 0V PNR data with the best fit corresponding to a uniform magnetization profile. (b) Best fit to the same data using a power law model. Error bars correspond to +/- 1 standard deviation.

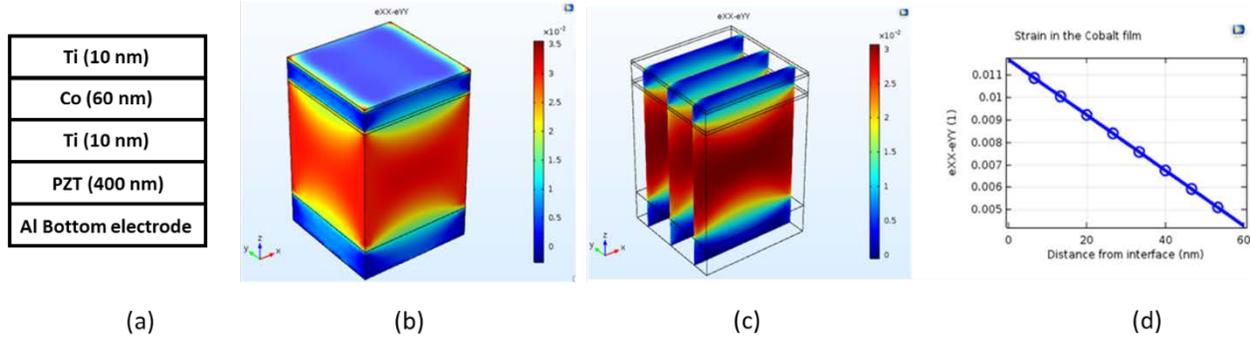

**Figure 8:** (a) Structure for the COMSOL model, (b) strain profile of the entire structure, (c) sliced strain profile, and (d) strain transfer through the center of the Co layer.

Due to computational limitations, the lateral dimensions were modeled to be 400 nm × 400 nm and the thickness of the piezoelectric layer was modeled to be 400 nm instead of the 0.5mm thickness of the substrates used in the experiments. A biaxial strain was applied to the PZT layer (2% tensile stress along x and 1% compressive stress along y) with the bottom surface of the bottom Al electrode assumed to be rigid. The corresponding strain transfer profile is shown for the entire modeled structure in Fig. 8(b) and a sliced version provides a better look at the strain transfer profile inside the structure as shown in Fig. 8(c). Fig 8(d) shows the strain transfer through the center of the Co layer as a function of distance from the Ti/Co interface nearest to the piezoelectric layer. The strain transfer changes monotonically with distance. If this strain transfer is responsible for the non-uniformity in the magnetization, it is reasonable to assume that magnetization rotation and magnitude should also change monotonically with depth. Thus, we propose that a monotonic function is appropriate for modeling of the neutron data. For simplicity, we chose a power law, where the endpoint magnetization magnitudes, rotation angles, and the power law exponent were treated as fitting parameters. This scheme results in a much better fit to the data, as shown in Fig. 7(b). The best-fit power law profiles for near-remanence and near-coercivity data are shown in Figure 9. Panel (a) shows the magnitude of the magnetization M, while panel (b) shows the rotation angle of the magnetization $\phi_M$, expressed in terms of the angle between the magnetization vector and the applied magnetic field axis. Note that for the near-remenant conditions, the component of magnetization along the H axis points the same direction as H, while for the near-coercivity conditions it points opposite. The best-fit parameters are shown in Table 1.

**Table 1:** Best-fit parameters for the power law models. Error bars correspond to 2 standard deviations.

| magnetic state | V (V) | bot M (kA m$^{-1}$) | top M (kA m$^{-1}$) | bot $\phi_M$ (deg) | top $\phi_M$ (deg) | Exponent |
|---|---|---|---|---|---|---|
| Remanent | 0 | 1132 ± 9 | 1206 ± 31 | 24.1 ± 0.9 | 5.5 ± 1.5 | 8.0 ± 1.7 |
| Remanent | 400 | 1174 ± 9 | 1187 ± 29 | 23.0 ± 0.9 | 5.2 ± 1.5 | 7.7 ± 1.7 |
| Coercive | 0 | 1248 ± 17 | 737 ± 22 | 33.3 ± 1.6 | 20.5 ± 0.6 | 1.3 ± 0.1 |
| Coercive | 400 | 1267 ± 16 | 788 ± 22 | 29.0 ± 1.4 | 16.4 ± 0.7 | 1.5 ± 0.1 |

The models for the near remanence (1 mT) and near coercive field (10 mT) are discussed in the following sections.

*Near Remanence (1 mT external magnetic field at 0V and 400V)* – In the 0V case, although there is no voltage induced stress, a residual stress is present in the Co film from poling of the underlying piezoelectric PMN-PT substrate. The top and bottom layers are rotated by 5.5°, and 24.1° (see Table I) from the saturation field direction. Among all the layers, the bottom layer is rotated by the highest amount. This rotation is the result of the residual stress present in the Co film. The effect of this stress is the most on the bottom layer (closest to the PMN-PT substrate), which is what the fitting suggests with the bottom layer showing the highest amount of rotation. The Co film is more relaxed (less stress) as we go up, which is also supported by a monotonically decreasing angle of magnetization rotation from the bottom towards the top. When an electrical voltage of 400V is applied across the sample thickness, the resulting electric field generates a stress in the substrate that is transferred to the Co film. The magnetizations of the Co layers respond to this voltage generated stress by rotating towards the applied magnetic field direction. However, the resulting rotation is very small (<1°). The rotation of magnetization towards the saturation field direction due to voltage induced stress matches the observations from section III.A. At both 0V and 400V, there is a clear difference in the amount of rotation between layers, confirming depth dependent variation in the layer magnetizations.

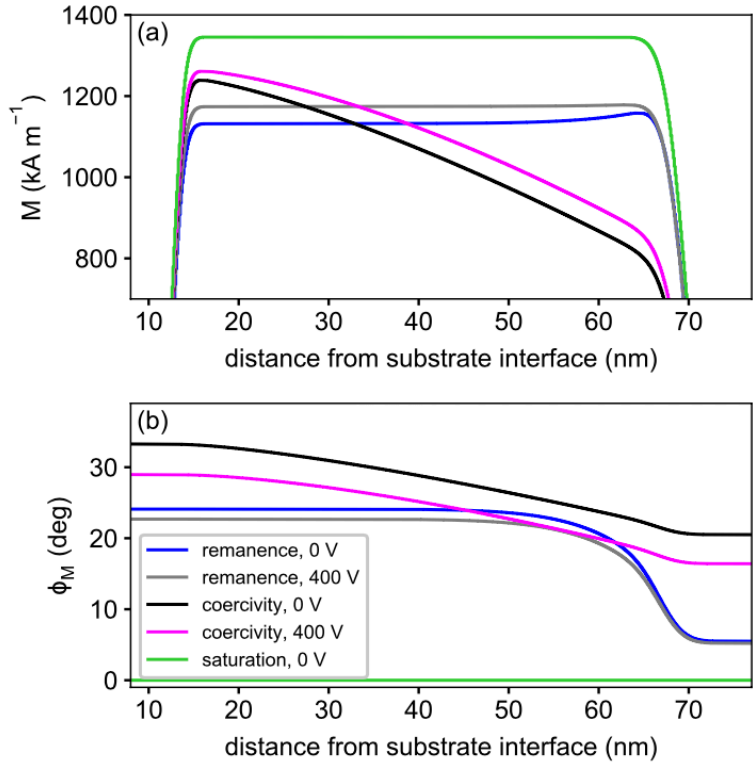

**Figure 9:** (a) Magnetization magnitude and (b) magnetization rotation profiles corresponding to the best-fit power law models for the near-remanent and near-coercivity PNR data. The saturation profiles are shown for comparison.

*Near Coercive field (10 mT external magnetic field at 0V and 400V)* – The measurements and fitting near coercive field is quite similar to the measurements at remanence. The difference in measurement conditions is that the sample was first saturated in the negative direction (-700 mT applied magnetic field) which was then increased to 10 mT. Again, similar to the near remanence case, a power law fit is applied. The fitted model and best fit parameters are shown in Figure 9 and Table I. We can see that both the magnetization orientation and the average magnetization has a depth dependent profile. In the 0V case, the effect of residual stress is the most in the bottom layer (deviation of 33.3°) and least in the top layer (deviation of 20.5°). For a small change in magnetic anisotropy, we can expect a larger change in the magnetization near the coercive field compared to the change near remanence (1mT, decreasing the field from positive saturation). So, the effect of voltage induced stress is expected to be larger near the coercive field compared to remanence. This is exactly what we see from the data in table I. Here, the bottom layer is rotated by ~4° as a result of the voltage induced stress whereas the rotation was ~1° in the near remanence case.

Now if we look at the average magnetization profile, we can see that unlike the remanence case, there is a clear depth dependence. The average magnetization is highest at the bottom layer and lowest at the top. A larger stress in the bottom layer provides a higher anisotropy which possibly results in a magnetic domain distribution where the magnetization directions are more likely to point along the same direction, resulting in a larger average magnetization. The top layer experiences a lower stress induced anisotropy possibly allowing the magnetization directions of the domains to have a larger distribution, resulting in a smaller average magnetization. So, the layers exhibits a monotonically decreasing average magnetization from the interface towards the surface of the Co layer as a result of the similarly decreasing stress induced anisotropy. At 400V, the voltage induced strain reinforces the magnetic anisotropy along the neutron polarization, we see a corresponding increase in the average magnetization of all layers.

## IV. CONCLUSION

In conclusion, the reflectivity data and the subsequently fitted models clearly show that the voltage induced stress, although very small, has a measurable effect on the magnetization of the Co thin film. The strain induced anisotropy increases the incoherency in the magnetization rotation process as is evident from the "hysteresis-like" measurements performed in section III.B. The most important observation in this study is the non-uniform magnetization rotation and average magnetization along the depth of the thin film, a clear indication of depth dependent stress anisotropy in these structures. This study confirms magnetization variation along the thickness of a magnetostrictive thin film which appears to be related to relaxation in strain transfer from the piezoelectric substrate to the magnetostrictive layer as we go upwards from the piezoelectric-magnetostrictive heterostructure interface towards the surface of the thin film. This strain variation will possibly be more serious in a patterned 100~200 nm lateral dimension nanostructure even if it is 10 nm thick, which is typically the size for strain mediated nanomagnetic devices. The results presented in this paper is repeatable across samples (The PNR reflectivity data and the corresponding fitted models for a similar but different sample can be found in the supplement).

## ACKNOWLEDGEMENT

M.M. A and J.A. were funded by NSF grants CCF Career 1253370 and ECCS 1609303. The Co films were deposited at Center for Nanoscale Science and Technology (CNST) Nanofab at NIST, Gaithersburg.

SUPPLEMENT

# Polarized Neutron Reflectometry Study of Depth Dependent Magnetization Variation in Co Thin Films on a PMN-PT Substrate

Md Mamun Al-Rashid[1,2,*], Alexander Grutter[3,+], Brian Kirby[3,×] & Jayasimha Atulasimha[1,2,∥]

[1]Mechanical and Nuclear Engineering, Virginia Commonwealth University, Richmond, VA.
[2]Electrical and Computer Engineering, Virginia Commonwealth University, Richmond, VA.
[3]NIST Center for Neutron Research, NIST, Gaithersburg, MD.

[*]alrashidmm@vcu.edu, [+]alexander.grutter@nist.gov, [×]brian.kirby@nist.gov, [∥]jatulasimha@vcu.edu

In the main paper, a power law fit was used to explain the polarized neutron reflectometry (PNR) data, which is motivated by the monotonic strain relaxation from the magnetostrictive thin film - piezoelectric substrate interface to the free surface of the magnetostrictive film based on finite element analysis simulations. This supplement presents a different model where the magnetostrictive Co film is divided into three consecutive layers of the same sample (referred to as sample 1 from here onward). This supplement also presents a three layer model for the PNR data from a second sample (sample 2). During the PNR measurement, sample 2 was oriented such that the voltage induced strain rotates the magnetization away from the applied field H direction. However, sample 2 also exhibits non-uniform depth dependent magnetization profile both at zero and non-zero applied electric field in the PMN-PT substrate. Findings from sample 2 reinforces the conclusion drawn from sample 1.

## I. Three layer model – Sample 1

In this modeling, an approach similar to the power law fitting in the main paper was used. The model was first fitted with data obtained at saturation (700 mT) and the nuclear profile obtained from this fit was used for all subsequent modeling. The average magnetization M and magnetization rotation away from the applied field direction are shown in Fig. S1.a and S1.b respectively. The best fit parameters are summarized in Table S1. The three layer fit shows similar behavior to that of the power law fit. Both the average magnetization and magnetization rotation are non-uniform along the depth of the Co film. The bottom layer is the furthest away from the saturation direction for all cases. An applied voltage of 400V across the piezoelectric PMN-PT substrate produces ~3° rotation near remanence and ~6° rotation near coercive fields on the bottom Co layer.

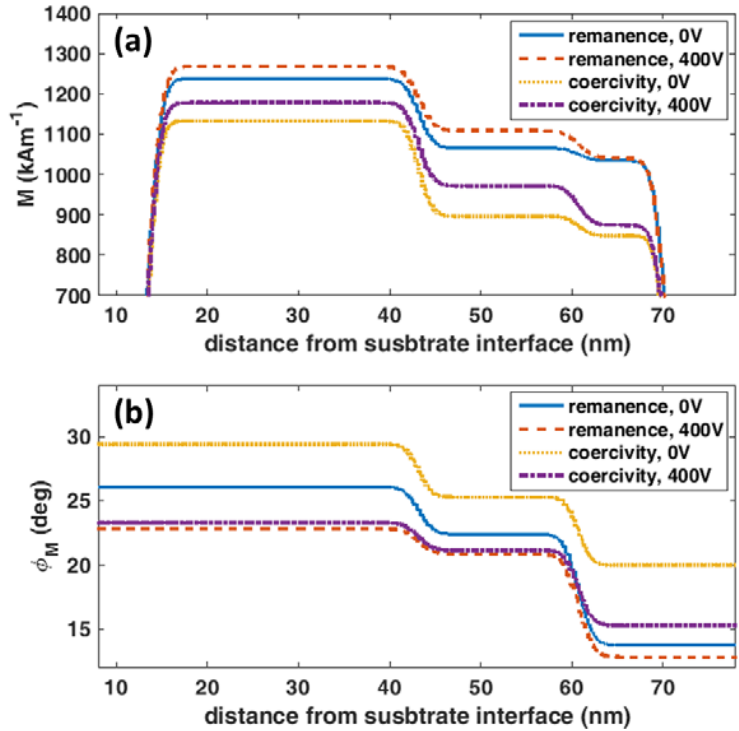

**Figure S1:** (a) Magnetization magnitude and (b) magnetization rotation profiles corresponding to the best-fit 3-layer models for the near-remanent and near-coercivity PNR data.

**Table S1:** Sample 1 - Best fit parameters

| Layer | Magnetization rotation (degrees) | | Average Magnetization M (kA/m) | | Thickness (Å) |
|---|---|---|---|---|---|
| | 0V | 400V | 0V | 400V | |
| **Near Remanence (1 mT)** | | | | | |
| **Top Layer** | 13.8° | 12.8° | 3.013 | 3.03 | 100 |
| **Mid Layer** | 22.4° | 20.8° | 3.102 | 3.227 | 175.4 |
| **Bottom Layer** | 26.1° | 22.8° | 3.604 | 3.692 | 300 |
| **Near Coercive Field (10 mT)** | | | | | |
| **Top Layer** | 20° | 15.3° | 2.47 | 2.547 | 100 |
| **Mid Layer** | 25.3° | 20.1° | 2.61 | 2.828 | 175.4 |
| **Bottom Layer** | 29.4° | 23.3° | 3.298 | 3.428 | 300 |

## II.     Repeatability: Sample 2 – Three-layer model

Similar measurements have been performed on a second sample (PMN-PT/Ta $_{(10nm)}$/Co $_{(60nm)}$). The only difference between the two samples is that sample 2 did not have the Ta capping layer. Further the stress axis of this sample was oriented differently w.r.t the applied field (H) direction compared to the sample-1 discussed in the main paper. Here the voltage induced stress produces a magnetization rotation away from the field (H) direction. Again, reflectivity measurements were performed at saturation (700mT, 0V) to estimate the nuclear parameters of the sample, followed by reflectivity measurements near remanence (1 mT) under both 0V and 400V applied voltage across the PMN-PT substrate. A three layer model similar to the previous section has been utilized to fit the PNR data for these measurements. The depth dependent magnetization rotation is shown in Fig. S2 and summarized in Table S2.

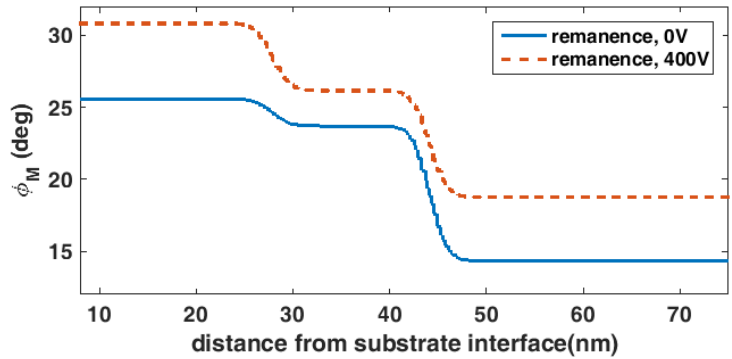

**Table S-II:** Fitted model parameters at remanence.

|  | Magnitude of rotation from 270° | |
| --- | --- | --- |
|  | 0V | 400V |
| **Top Layer** | 14.3° | 18.8° |
| **Mid Layer** | 23.7° | 26.2° |
| **Bottom Layer** | 25.6° | 30.8° |

Figure S2 and table S-II also show a depth dependent magnetization orientation profile both at zero and non-zero electric field similar to sample 1. In the 0V case, the effect of the residual stress is the most significant on the bottom layer and reduces as we go to the upper layers. In the 400V case, the voltage induced stress forces the magnetization further away from the applied field direction, opposite to what we see for sample 1. This is because the orientations of the two samples during loading were different, causing the magnetization to move toward the applied field direction for sample 1 and away for sample 2. This nevertheless shows that the observed variation in depth dependent magnetization orientation and strain transfer is repeatable over multiple samples. In this sample (sample 2), the voltage induced strain causes the magnetization to rotate away from the saturation direction, whereas in the sample (sample 1) presented in the main paper the voltage induced strain causes the magnetization to rotate towards the saturation direction. This is shown in Fig. S3.

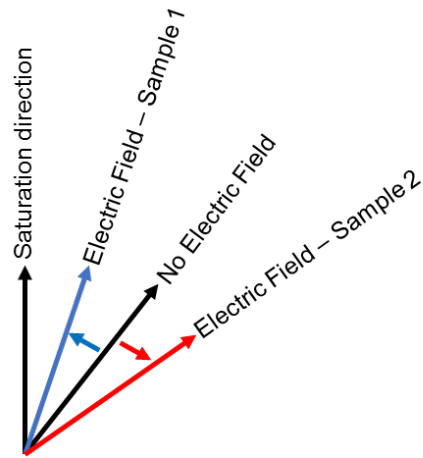

**Figure S3:** Direction of magnetization rotation under voltage induce strain.